\documentclass[]{article}
\usepackage{graphicx}                    
\usepackage[square]{natbib}
\begin{document}
\title{Radio Bursts in the Active period January 2005}

\author{C. Bouratzis, P. Preka--Papadema, A. Hillaris, X. Moussas,C. Caroubalos\\{\em{University of Athens, 15783 Athens, Greece}}\\
V. Petoussis, P. Tsitsipis, A. Kontogeorgos\\{\em{Technological Education Institute of Lamia, Lamia, Greece}}\\}
\maketitle
\begin{abstract}
We present complex radio bursts recorded by the radiospectrograph ARTEMIS-IV in the active period of January 2005. The wide spectral coverage of this recorder, in the 650-20 MHz range, permits an analysis of the radio bursts from the base of the Solar Corona to 2 Solar Radii; it thus facilitates the association of radio activity with other types of solar energetic phenomena. Furthermore the 
ARTEMIS-IV\footnote{Appareil de Routine pour le Traitement et l' Enregistrement Magnetique de l' Information Spectral}, 
high time resolution (1/100 sec)  in the 450--270 MHz range, makes possible the detection and analysis of the fine structure which most of the major radio events exhibit.
\end{abstract}

\section{Instrumentation \& Observations}
\begin{figure}[h]
\includegraphics[width=\textwidth,height=5cm]{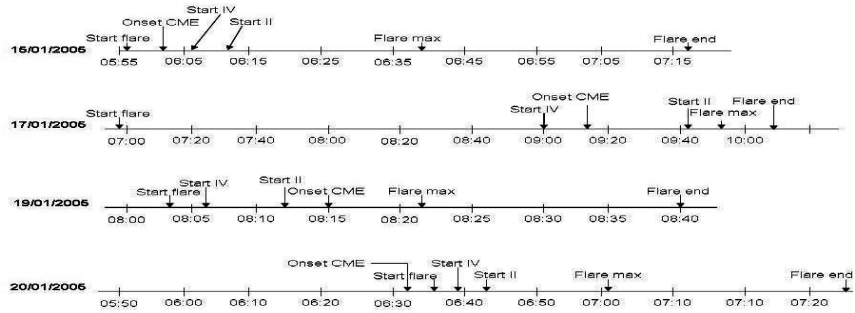}
  \caption{Schematic Presentation of the Events Time Sequence.}
\label{timeseries}
\end{figure}
The Artemis--IV solar radio--spectrograph (\cite{Caroubalos01a}, \cite{Caroubalos06}, \cite{Kontogeorgos})
covers the metric--decametric range  with two receivers operating in parallel: A sweep frequency analyser (ASG) 
covering the 650--20 MHz range at 10 samples/sec and a high sensitivity acousto--optical analyser (SAO), 
which covers the 270--450 MHz range with a time resolution of 100 samples/sec.
\begin{figure}
\begin{minipage}[t]{6cm}
\begin{center}
\includegraphics[width=6cm,height=4cm]{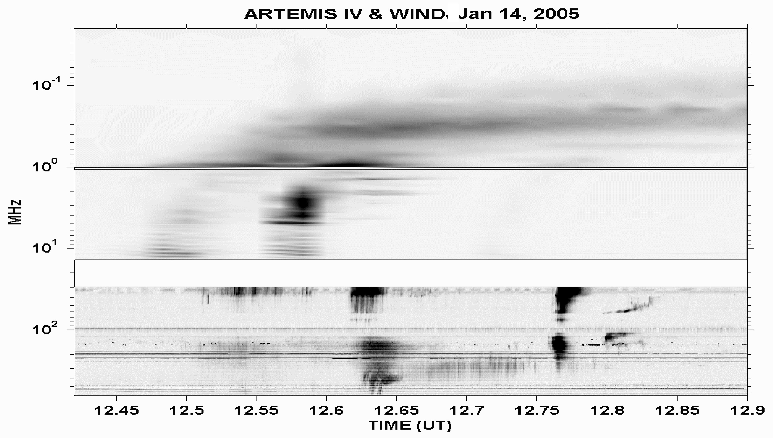}
\end{center}
\end{minipage}
\hfill
\begin{minipage}[t]{6cm}
\begin{center}
\includegraphics[width=6cm,height=4cm]{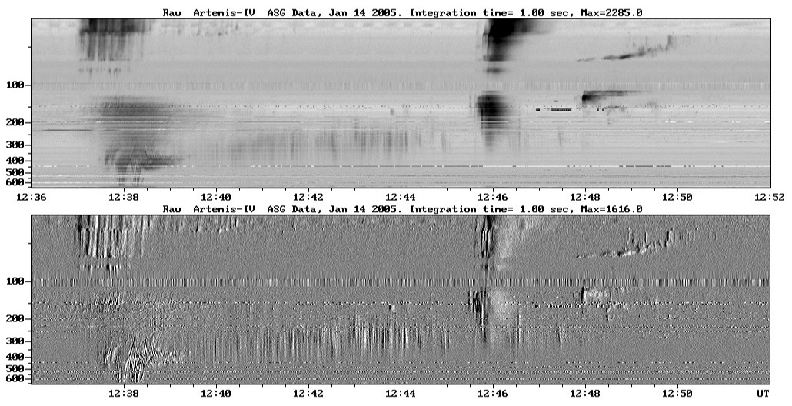}
\end{center}
\end{minipage}
 \includegraphics[width=\textwidth,height=3cm]{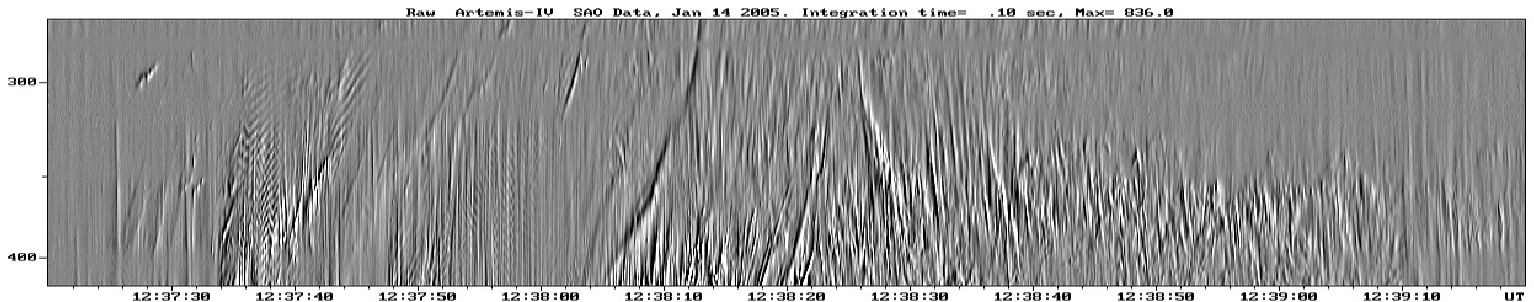}
\begin{minipage}[t]{6cm}
\begin{center}
\includegraphics[width=6cm,height=4cm]{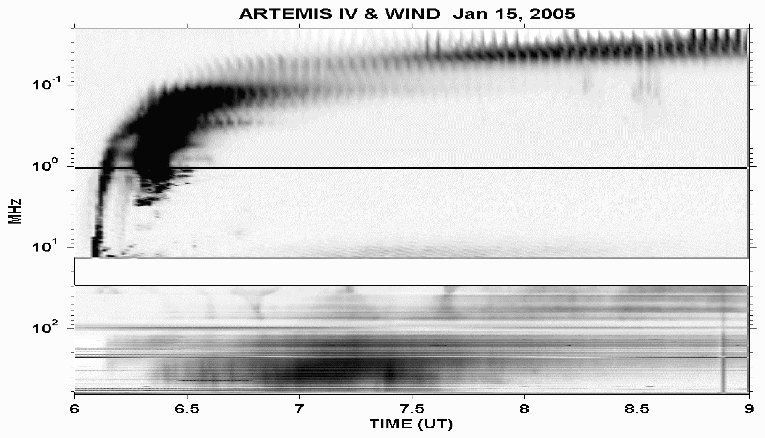}
\end{center}
\end{minipage}
\hfill
\begin{minipage}[t]{6cm}
\begin{center}
\includegraphics[width=6cm,height=4cm]{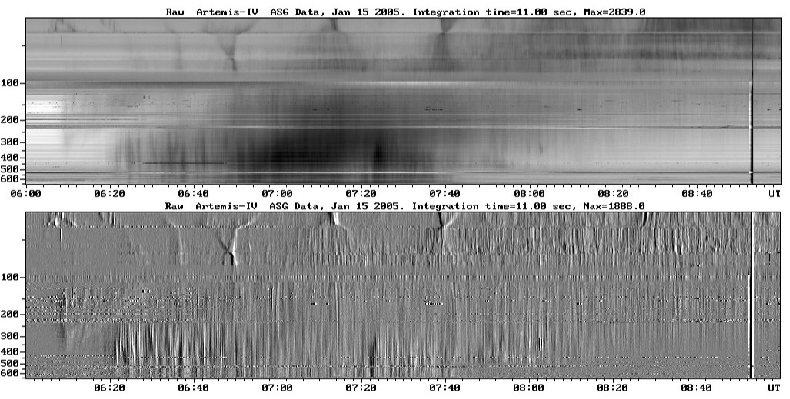}
\end{center}
\end{minipage}
 \includegraphics[width=\textwidth,height=3cm]{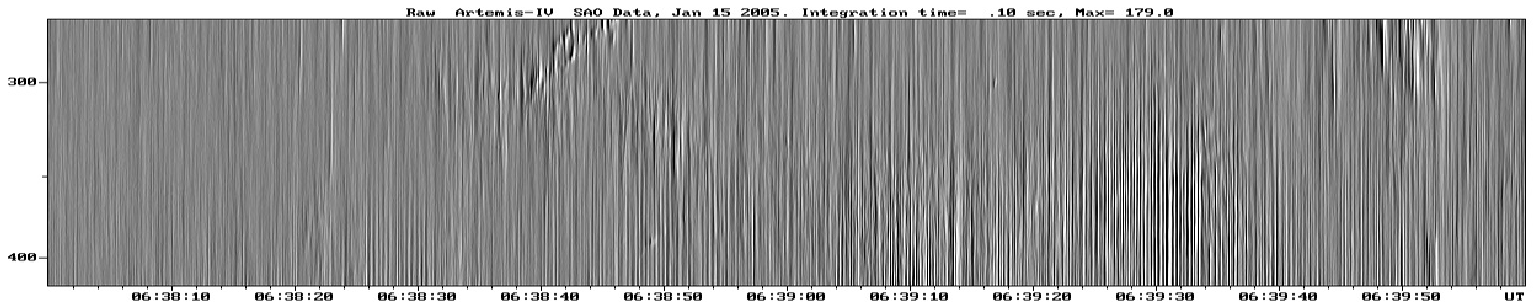}
  \caption{UPPER PANEL: LEFT: Combined ARTEMIS-IV and WIND--WAVES spectra of the January 14, 2005 event.
RIGHT: ARTEMIS--IV ASG Spectra in the frequency range 20--650 MHz; 
Intensity (top)and its time derivative (bottom), enhancing fine temporal details.
SECOND FROM TOP: Fibers, Zebra, Pulsations and \emph{Lace Bursts} (cf. \cite{Jiricka} from the same event (Differential
Spectrum).)
THIRD PANEL: Same as UPPER PANEL but for the event of January 15, 2005.
LAST PANEL:Narrow Band Spikes \& Pulsations from the same event (Differential Spectrum).}
 \label{05114_01}
 \end{figure}
 In the active period 14--20 January 2005 the ARTEMIS--IV observed five major events
January 14 (12:36 UT cf. figure \ref{05114_01} Top Two panels), 
January 15 (06:06 UT cf. figure \ref{05114_01} Bottom Two panels), 
January 17 (09:00 UT cf. figure \ref{05117_01} Top Two panels), 
January 19 (08:05 UT cf. figure \ref{05117_01} Bottom Two panels),
January 20 (06:39 UT cf. figure \ref{05120_01})
Combined with data from WIND--WAVES (\cite{Bougeret}), these observations provide a complete view of
the radio emission induced by shock waves and electron beams from the low corona to about 1 A.U. 

These recordings were supplemented with CME data from the LASCO lists 
on line\footnote{http://cdaw.gsfc.nasa.gov/CME list} (\cite{Yashiro}) and 
SXR (GOES) reports from SGD\footnote{http//www.sel.noaa.gov/ftpmenu/indices}; the 20 January 2005 CME kinetic data
were from \cite{Gopal}. A brief overview of the five events is presented in the Table; a schematic presentation of the corresponding time
sequences is displayed in figure \ref{timeseries}.
\begin{figure}
\begin{minipage}[t]{6cm}
\begin{center}
\includegraphics[width=6cm,height=4cm]{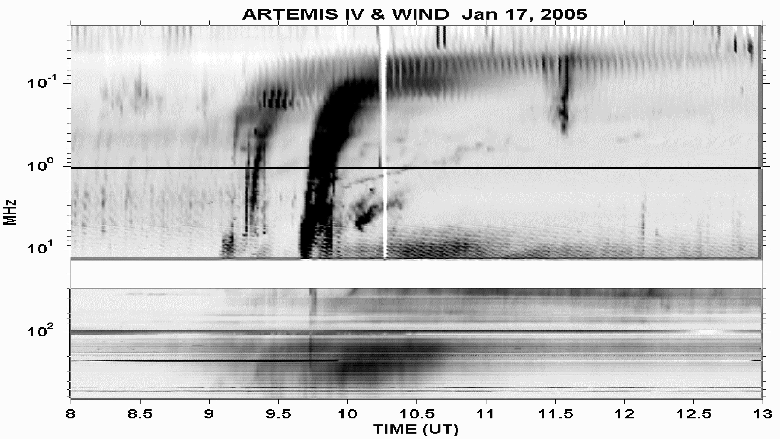}
\end{center}
\end{minipage}
\hfill
\begin{minipage}[t]{6cm}
\begin{center}
\includegraphics[width=6cm,height=4cm]{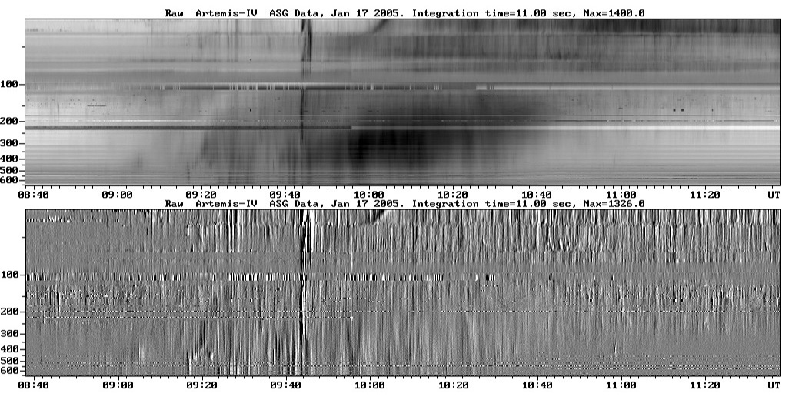}
\end{center}
\end{minipage}
 \includegraphics[width=\textwidth,height=3cm]{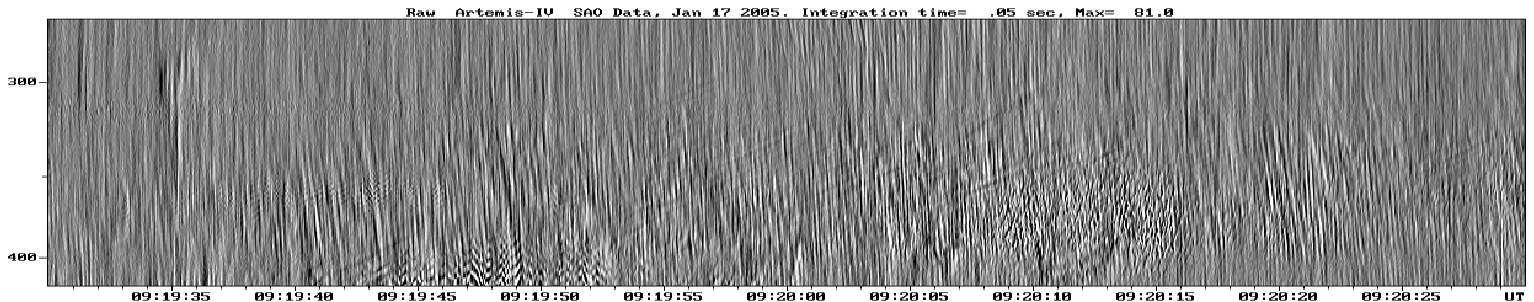}
\begin{minipage}[t]{6cm}
\begin{center}
\includegraphics[width=6cm,height=4cm]{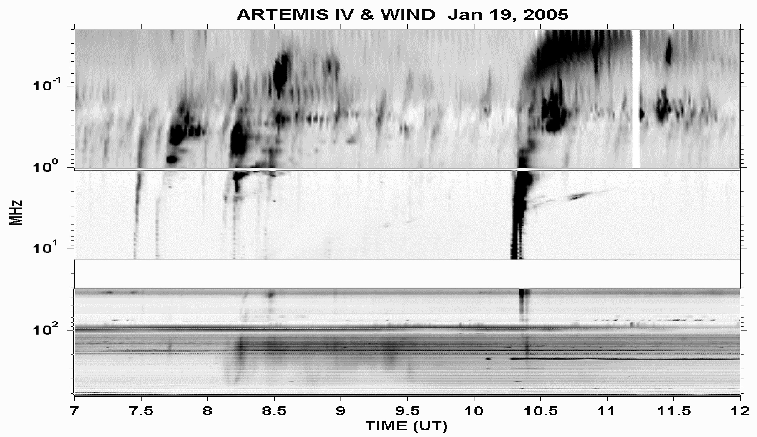}
\end{center}
\end{minipage}
\hfill
\begin{minipage}[t]{6cm}
\begin{center}
\includegraphics[width=\textwidth,height=4cm]{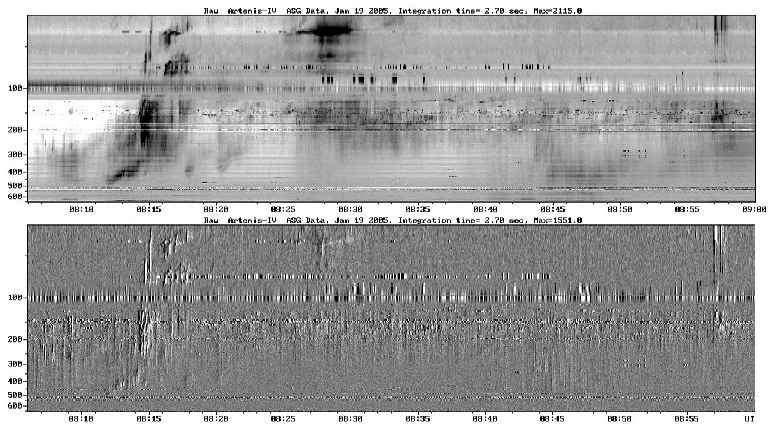}
\end{center}
\end{minipage}
 \includegraphics[width=\textwidth,height=3cm]{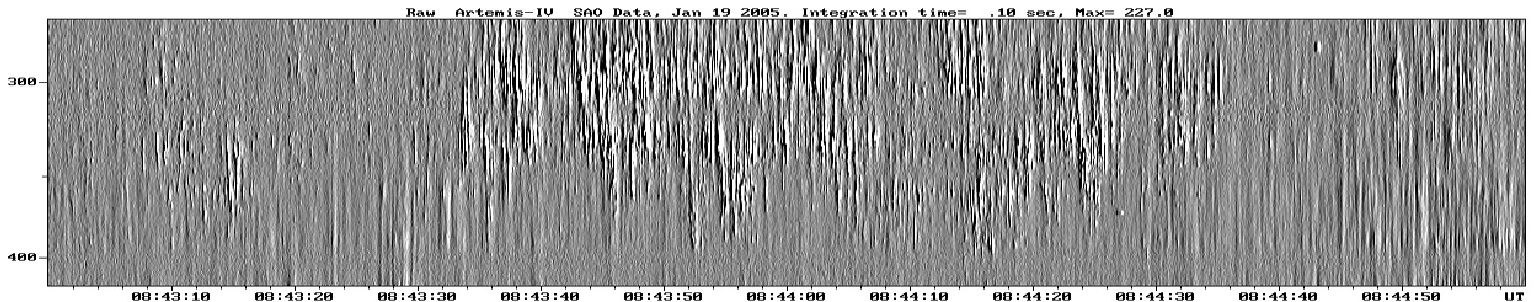}
  \caption{UPPER PANEL: LEFT: Same as figure \ref{05114_01} but for the events of January 17, 2005.
SECOND FROM TOP: Fibers, Zebra, Pulsations and a Narrow band spike group from the same event (Differential
Spectrum).)
THIRD PANEL: Same as UPPER PANEL but for the event of of January 19, 2005.
LAST PANEL:Narrow Band Spikes \& Pulsations from the same event (Differential Spectrum).}
\label{05117_01}
 \end{figure}
\begin{figure}
\begin{minipage}[t]{6cm}
\begin{center}
\includegraphics[width=6cm,height=4cm]{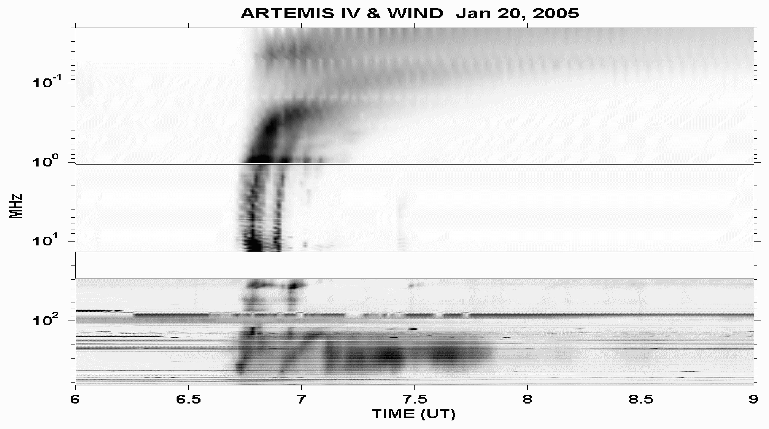}
\end{center}
\end{minipage}
\hfill
\begin{minipage}[t]{6cm}
\begin{center}
\includegraphics[width=6cm,height=4cm]{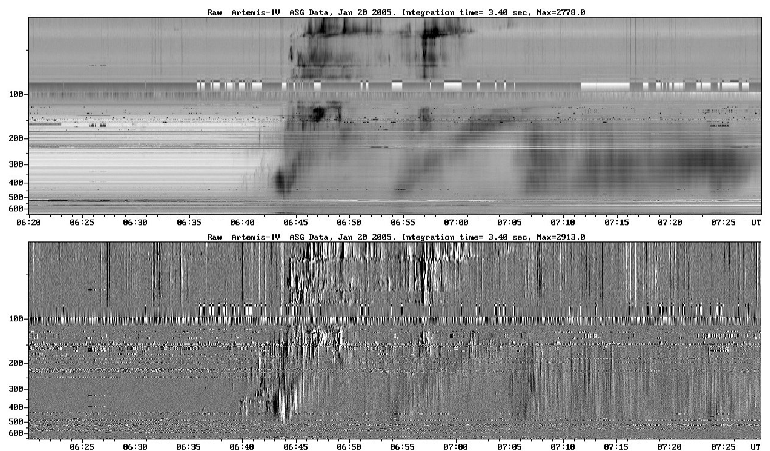}
\end{center}
\end{minipage}
 \includegraphics[width=\textwidth,height=3cm]{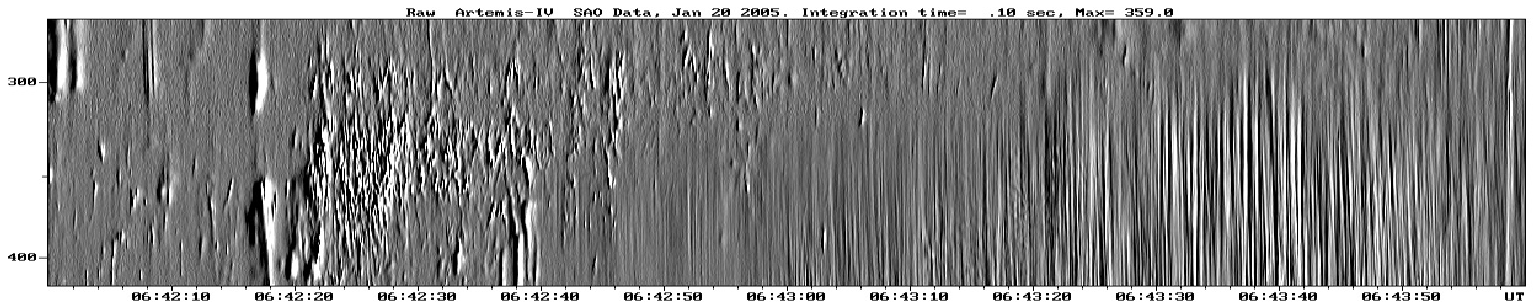}
 \caption{UPPER PANEL: Same as figures \ref{05114_01} \& \ref{05117_01} but for the event of January 20, 2005.
LOWER PANEL: Narrow Band Type III, Spikes  \& Pulsations from the same event (Differential Spectrum).}
\label{05120_01}
\end{figure}
The high sensitivity and time resolution of the SAO facilitated an examination on fine structure 
within the three studied periods; all exhibited rich fine structure embedded in the Type--IV continua.
In our analysis, the continuum background is removed by the use  of high--pass filtering on the
dynamic spectra (differential spectra in this case).

We present certain examples, which are divided according to a published morphological 
classification scheme (\cite{Jiricka}) based on Ondrejov Radiospectrograph recordings in the
0.8--2.0 GHz range. In our recordings we have detected:
\begin{table}[]
\begin{tabular}{{lllllll}} %
\hline\hline
\textbf{Event}	&\textbf{Start}	   & \textbf{AR}&\textbf{Freq. Range} &\textbf{Class}&\textbf{Speed}&\textbf{Accel}\\
			&\textbf{UT}         & 		   &\textbf{MHz} 	  &		  &\textbf{Km/sec}&\textbf{$Km/sec^2$}\\
\hline
\textbf{14 Jan 05}&			   & 		    &			        &		     &              &             \\
Type II		&	12:48 	   &	718       &			30--70  &		     & 1190         &      -      \\
SXR			&	12:33 	   &	718       &			        &	C4.6	     &              &             \\
\hline
\textbf{15 Jan 05}&			   & 		    &			        &		     &              &             \\
Type II		&	06:08 	   &	720       &		100--400  	  &		     & 690          &             \\
Type IV		&	06:15 	   &	720       &		20--650  	  &		     &              &             \\
SXR			&	05:54 	   &	720       &			        &	M8.6	     &              &             \\
CME (lift off)	&	06:00 	   & 		    &			        &		     &  2049        &   -30.7      \\
\hline
\textbf{17 Jan 05}&			   & 		    &			        &		     &              &             \\
Type II		&	09:43 	   &	720       &		150--500  	  &		     & 3256         &             \\
Type IV		&	09:16 	   &	720       &		20--400  	  &		     &              &             \\
SXR			&	06:59 	   &	720       &			        &	X3.8	     &              &             \\
CME (lift off)	&	09:15 	   & 		    &			        &		     & 2094         &    -118.0    \\
\hline
\textbf{19 Jan 05}&			   & 		    &			        &		     &              &             \\
Type II		&	08:12 	   &	720       &		20--500  	  &		     & 1624         &             \\
Type IV		&	08:06 	   &	720       &		150--400  	  &		     &              &             \\
SXR			&	08:03 	   &	720       &			        &	X1.3	     &              &             \\
CME (lift off)	&	08:15 	   & 		    &			        &		     & 2020         &   -43.8     \\
\hline
\textbf{20 Jan 05}&			   & 		    &			        &		     &              &             \\
Type II		&	06:43 	   &	720       &		20--500  	  &		     &  2294        &             \\
Type IV		&	06:39 	   &	720       &		150--450  	  &		     &              &             \\
SXR			&	06:36 	   &	720       &			        &	X7.1	     &              &             \\
CME (lift off)	&	06:33 	   & 		    &			        &		     &  3675        &   ---        \\
\hline\hline
\end{tabular}
\caption{Overview of the Events 14--20 January 2005}
\label{tab}
\end{table}
\begin{itemize}
 \item{Narrow Band Spikes (figures \ref{05114_01}, \ref{05117_01}, \ref{05120_01})}
\item{Fibers (figures \ref{05117_01},\ref{05114_01})}
\item{Narrowband Type III Bursts (figure \ref{05120_01})}
\item{Laces (figure \ref{05114_01})}
\item{Zebra patterns (figure \ref{05117_01})}
\end{itemize}

\section{DISCUSSION \& CONCLUSIONS}
The ARTEMIS--IV radio--spectrograph, operating in the range of 650--20 MHz, observed 5 complex events during
the super--active period 14--20 January 2005. WIND-WAVES data complement nicely the ARTEMIS data and trace the radio 
emission from the middle corona all the way to almost 1 A.U. The high resolution SAO recordings on the other hand
reveal a variety of fine structure which, almost, matches the comprehensive Ondrejov Catalogue 
(\cite{Jiricka}). This last, although it refers to the spectral range 0.8--2 GHz, seems to produce
similar fine structure with the metric range. 
\\
\\
{\em{The LASCO CME catalogue is generated and maintained by the Center for Solar Physics
and Space Weather, The Catholic University of America in cooperation
with the Naval Research Laboratory and NASA. SOHO is a project of international cooperation between ESA and NASA. 
The Solar Geophysical Data Catalogue is compiled and maintained by the US Department of Commerce.}}

\end{document}